\documentstyle[11pt,newpasp,twoside,epsf]{article}

\begin{document}
\title{E+A Galaxies in the near-IR: Field and Clusters}
\author{Gaspar Galaz}
\affil{Carnegie Institution of Washington. Las Campanas Observatory. Casilla
601. La Serena. Chile. E-mail: {\tt gaspar@azul.lco.cl}}

\begin{abstract}
I present near-IR photometry of a selected sample of E+A
galaxies observed in the southern hemisphere. The sample includes 50 galaxies from 
nearby clusters ($z \sim 0.05$) and distant clusters ($z \sim 0.31$)
as well as E+A galaxies from the field ($z \sim 0.15$).
I have also observed 13 normal early-type galaxies from the field and from clusters,
to be compared with the E+A sample. 
The photometry includes $J$, $H$ and $K_s$ apparent magnitudes and colors. 
I investigate systematic properties of the E+A sample as a 
function of their environment, contrasting the observed colors with
spectrophotometric models of galaxy evolution.  
\end{abstract}

\section{Introduction: what is an E+A galaxy?}

Most of the E+A galaxies present mid- to early-type
morphologies, as it was already shown by some authors ([1], [2]). 
A little fraction of them have late Hubble types (but see [3]). However, 
their spectra is peculiar: they do not have emission lines, representative of
an ongoing star formation, but they have strong Balmer
absorption lines, representative of a young population (A and B spectral types),
and also strong Mg b $\lambda 5175$, Ca H \& K 
$\lambda 3934$, $\lambda 3968$ and Fe $\lambda5270$ lines, indicating 
that they have a rich population of G, K and M spectral types.
This young population suggests that the E+A galaxies are 1 Gyr to 4 Gyr
old. Are there other peculiar signatures in the spectra of the E+A galaxies? 
In particular, are their late-type star populations and their AGB population 
also different to those of normal galaxies? If so, we can conclude
that the E+A phenomenon also involves changes in their older population.
In that case, models trying to explain the nature of the E+A galaxies 
should also fit the signatures observed in the old stellar content.

\section{Near-IR photometry of E+A galaxies}

\begin{figure}
\plotfiddle{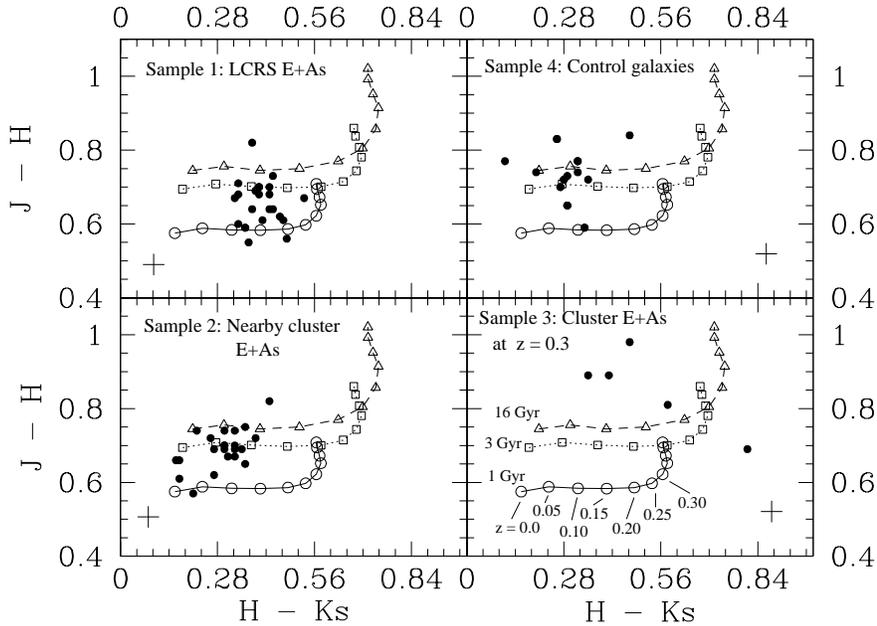}{7cm}{-90}{45}{45}{-180}{255}
\caption{Observed near-IR colors (solid dots) for different samples of galaxies. 
Sample 1 are field E+A galaxies from the Las Campanas Redshift Survey [5], at 
$z \sim 0.15$. Sample 2 are E+A galaxies from nearby clusters [2]. Sample 3
are E+As from $z \sim 0.3$ clusters [4]. Sample 4 are normal nearby galaxies.
Open connected symbols represent corresponding colors of spectrophotometric models
of galaxy evolution (GISSEL96 [7]) at different redshifts. Different symbols correspond
to different ages after an instantaneous burst, considering 
solar metallicity, as indicated in the Figure.}
\label{colors}
\end{figure}

In order to investigate the above questions, I have started a 
program to observe southern E+A galaxies in the near-IR. 
All the observations are being carried on at Las Campanas Observatory (LCO),
using NICMOS3 HgCdTe arrays (256 $\times$ 256 pixels)
at both the 1-m Swope telescope (0.599 arcsec/pix, 2.5 arcmin$^2$ FOV), and
the 2.5-m du Pont telescope (0.42 arcsec/pix, 1.8 arcmin$^2$ FOV). 
All the observations are carried on under photometric conditions and 
seeing $\la 1.0$ arcsec, and include $J$, $H$ and $K_s$ 
imaging. The sample of galaxies includes E+As from nearby ($z \la 0.05$, [2])
and intermediate-redshift ($z \sim 0.3$, [4]) clusters, as well as E+As
located in the field [5], at $z \sim 0.15$.  
Photometry is performed on the calibrated images
using SExtractor [6]. {\em Total} apparent magnitudes and colors
are computed, and compared with spectrophotometric models of galaxy 
evolution generated using GISSEL96 [7] (see Figure \ref{colors}). 
Rest-frame colors are
extremely dependent on models, necessary to compute K-corrections.
Near-IR spectroscopy will be obtained soon, in order to have 
reliable K-corrections, allowing to derive robust interpretations from
rest-frame colors. 

\end{document}